\documentstyle[prl,twocolumn,aps,graphicx]{revtex}

\begin{document}
\title{Conductance of Distorted Carbon Nanotubes}
\author{Alain Rochefort$^*$, Fr\'ed\'eric Lesage$^{**}$, Dennis R. 
Salahub$^{*,\dagger}$, Phaedon Avouris$^\ddagger$}
\address{$^*$ Centre de recherche en calcul appliqu\'e (CERCA),
5160 Boul. D\'ecarie, bureau 400, Montr\'eal, (Qu\'ebec) Canada H3X-2H9.}
\address{$^{**}$ Centre de recherche math\'ematiques (CRM),
Universit\'e de Montr\'eal, C.P. 6128, Succ. Centre-Ville, 
Montr\'eal, (Qu\'ebec) Canada H3C-3J7.}
\address{$^\dagger$ D\'epartement de chimie, Universit\'e de Montr\'eal,
C.P. 6128, Succ. Centre-Ville, Montr\'eal, (Qu\'ebec) Canada H3C-3J7.}
\address{$^\ddagger$ IBM Research Division, T.J. Watson Research Center,
P.O. Box 218, Yorktown Heights, NY 10598, USA.}
\maketitle

\begin{abstract}
We have calculated the effects of structural distortions of armchair carbon
nanotubes on their electrical transport properties. We found that the bending
of the nanotubes decreases their transmission function in certain energy
ranges and leads to an increased electrical resistance. Electronic structure
calculations show that these energy ranges contain localized states with
significant $\sigma$-$\pi$ hybridization resulting from the increased
curvature produced by bending. Our calculations of the contact resistance
show that the large contact resistances observed for SWNTs are likely due to
the weak coupling of the NT to the metal in side bonded NT-metal
configurations.\\ 
\end{abstract}

Carbon  nanotubes (NTs) can be metallic or semiconducting. They have high
mechanical strength and good thermal conductivity \cite{dresselhaus},
properties that make them potential building blocks of a new, carbon-based,
nanoelectronic technology \cite{tans,bockrath,collins,martel}. Conduction in
defect-free NTs, especially at low temperatures, can be ballistic, thus
involving little energy dissipation within the NT \cite{white}.  Furthermore,
NTs are expected to behave like quasi one-dimensional systems (Q1D) with
quantized electrical resistance, which, for metallic armchair nanotubes at
low bias should be about 6 k$\Omega$ ($h/4e^2$). The experimentally observed
behavior is, however, quite different. The contact resistance of single-wall
nanotubes (SWNTs) with metal electrodes is generally quite high.
Furthermore, at low temperatures a localization of the
wavefunction in the nanotube segment contained between the metal electrodes
is observed that leads to Coulomb blockade phenomena \cite{bezryadin}. The
latter observation suggests that a barrier or bad-gap develops along the NT
near its contact with the metal.  In an effort to understand the origin of
these discrepancies we have used Green's function techniques to calculate the
effect of the modification of the NTs by bending on their electronic structure
and electric transport properties. We also investigated the effects
of the strength of the NT-metal pad interaction on the value of the contact
resistance.\\

Most discussions on the electronic structure of NTs assume perfect
cylindrical symmetry. The introduction of point defects such as vacancies
\cite{chico} or disorder \cite{white,anan} has been shown to lead to
significant modification of their electrical properties. Here we focus on the
effects of structural (axial) distortions on the transport properties of
armchair NTs. AFM experiments \cite{hertel1} and molecular mechanics
simulations \cite{hertel2} have shown that the van der Waals forces between
NTs and the substrate on which they are placed can lead to a significant
deformation of their structure. To maximize their adhesion energy the NTs
tend to follow the topography of the substrate \cite{hertel1,hertel2}. Thus,
for example, NTs bend to follow the curvature of the metal electrodes on
which they are deposited. When the strain due to bending exceeds a certain
limit, kinks develop in the nanotube structure \cite{hertel1,iijima,falvo}.
It is important to understand how these NT deformations affect the electrical
transport properties of the NTs. Could they be responsible for the low
temperature localization observed ? \cite{bezryadin} Early theoretical work
on this issue was based on tight-binding model involving only the
$\pi$-electrons of the NTs and accounted for the electronic structure changes
induced by bending through the changes in $\pi$-orbital overlap at neighboring
sites. This study concluded that bending distortions would have a negligible
effect on the electrical properties of the NTs \cite{kane}.  The
applicability of this approach is limited to weak distortions.  Experiments,
however, show that strong deformations and kink formation are common.  Under
such conditions, bending-induced  $\sigma$-$\pi$ mixing, which was not
considered before, becomes very important in strongly bent NTs
\cite{rochefort}.  In this work, the NT electronic structure is computed
using the extended H\"uckel method (EHM) \cite{yaehmop} that includes both
$s$ and $p$ valence electrons. We have previously \cite{rochefort2} shown
that EHM calculations on an armchair $(6,6)$ NT model (96 {\AA} long)
reproduce the electronic properties obtained with more sophisticated {\it
ab-initio} and band structure computations on NTs.
The approach we used in the computation
of the electrical properties is similar to that of Datta {\it et al.}
\cite{datta,dattamol}.\\

The conductance through a molecule or an NT cannot be easily computed; even
if the electronic structure of the free molecule or NT is known, the effect
of the contacts on it can be substantial \cite{lang} and needs to be taken
into account. Typically, there will be two (or more) leads connected to the
NT. We model the measurement system as shown in Figure 1a. The leads are
macroscopic gold pads that are coupled to the ends of the NT through matrix
elements between the Au surface atoms and the end carbon atoms of the NT. In
most experiments to date the NTs are laid on top of the metal pads. As we
discussed above, the NTs then tend to bend around the pads. Such bending
deformations are modelled in our calculations by introducing a single bend
placed at the center of the tube.\\

The electrical transport properties of a system can be described in terms of
the retarded Green's function \cite{datta,economou}. To evaluate the
conductance of the NT we need to compute the transmission function, $T(E)$,
from one contact to the other.  This can be done following the
Landauer-B\"uttiker formalism as described in \cite{datta}.  The key element
of this approach lies in the treatment of the infinite leads which are here
described by self-energies. We can write the Green's function in the form
of block matrices separating explicitly the molecular Hamiltonian. After some
simplification we obtain:
\begin{equation}
G_{NT}=\big[ ES_{NT}-H_{NT}-\Sigma_1-\Sigma_2 \big]^{-1}
\end{equation}
\noindent
where $S_{NT}$ and $H_{NT}$ are the overlap and the Hamiltonian matrices,
respectively, and $\Sigma_{1,2}$ are self-energy terms that describe the
effect of the leads. They have the form $\tau_{i}^\dagger g_{i} \tau_{i}$
with $g_{i}$ the Green's function of the individual leads
\cite{dattamol,papa} and $\tau_{i}$ is a matrix describing the interaction
between the NT and the leads.  The Hamiltonian and overlap matrices are
determined using EHM for the system Gold-NT-Gold. The transmission function,
$T(E)$, that is obtained from this Green's function is given by
\cite{datta}:

\begin{equation} 
T(E)=T_{21}=Tr [\Gamma_2 G_{NT} \Gamma_1 G_{NT}^\dagger ].  
\end{equation} 

In this formula, the matrices have the form:  

\begin{equation} 
\Gamma_{1,2}=i(\Sigma_{1,2}-\Sigma^\dagger_{1,2}).  
\end{equation} 

The summation over all conduction channels in the molecule allows the
evaluation of the resistance at the Fermi energy, $R=h/(2e^2T(E_F))$.
Transport in the presence of an applied potential is also computed.  The
differential conductance is computed in this case using the approximation
\cite{datta}:
\begin{equation}
\kappa (V)=\frac{\partial I}{\partial V} \approx \frac{2e^2}{h}
[\eta T(\mu_1)+(1-\eta )T(\mu_2)]
\end{equation} 
\noindent
with $\eta\in [0,1]$ describing the equilibration of the nanotube energy
levels with respect to the reservoirs \cite{dattamol}.  As a reference, we
use the $E_F$ obtained from EHM for individual nanotubes as the
zero of energy. The NT model used in our calculations is a $(6,6)$ carbon 
nanotube segment containing 948 carbon atoms. The bond distance between 
carbon atoms in non-deformed regions of the NT is fixed to that in graphite
1.42 {\AA}, leading to a tube length of 96 {\AA}. The building of deformed NTs 
using molecular mechanics minimization schemes \cite{mm3,tinker} has been
described in detail elsewhere \cite{rochefort}. The structures of the bent
NTs are shown in Figure 1b. The metallic contacts consist each of 22 gold
atoms in a (111) crystalline arrangement. The height of the NT over the gold
layer is 1.0 {\AA}, where the Au-C bond distances vary from 1.1 to
1.6 {\AA}. 

\begin{figure}[tbh]
{\Large \bf a}
\centerline{\includegraphics[width=6cm]{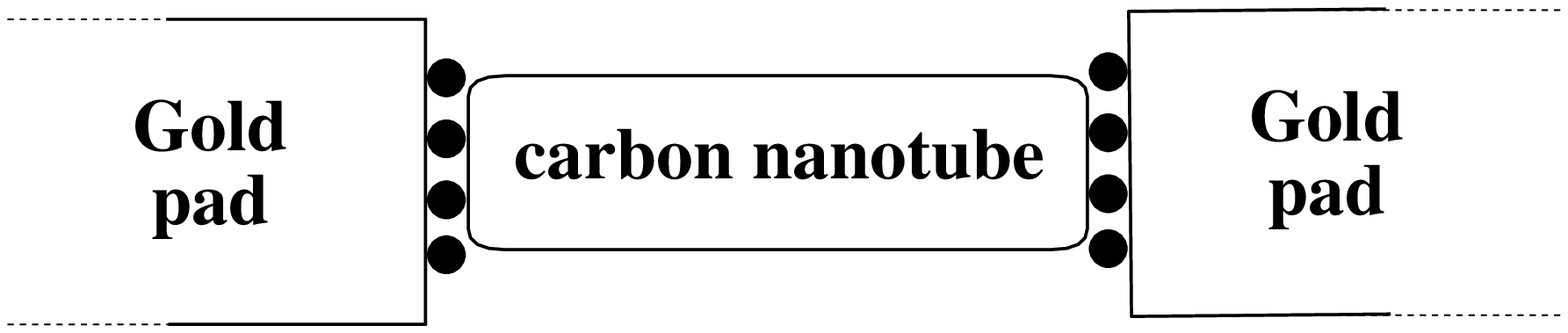}}

{\Large \bf b}
\centerline{\includegraphics[angle=90,width=8cm]{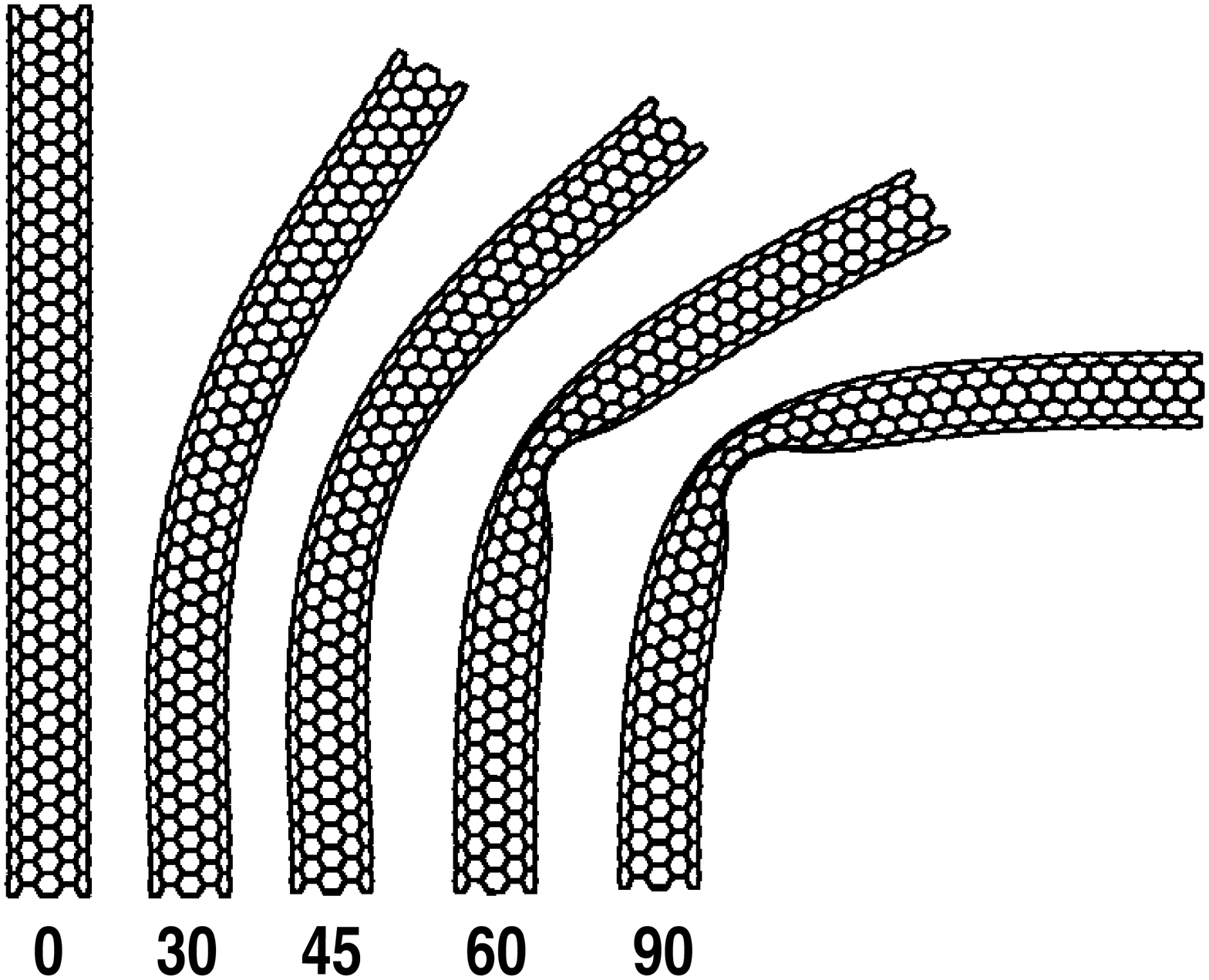}}
\caption{(a) Schematic of the model used in computation involving a
finite length carbon nanotube contained between infinite gold pads.
(b) Structures of the isolated bent nanotubes used in the computation.
The indices refer to the bending angles (in degrees).}
\end{figure}

In Figure 2 we present the computed transmission function $T(E)$ for the bent
tubes (note that $T(E)$ represents the sum of the transmission probabilities
over all contributing NT conduction channels). The upper-right of Figure 2
shows the raw transmission results obtained for the straight NT. The fast
oscillations of $T(E)$ are due to the discrete energy levels of the finite
segment of the carbon nanotubes used. For clarity, we will use smoothed curves in
the description of the results. At $E_F$, $T(E)$ is about 1.2, leading to a
resistance ($\approx$ 11 k$\Omega$ ) higher than expected for ballistic
transport ($\approx$ 6 k$\Omega$ for $T(E)$ = 2.0). This reduction in
transmission is due to the contribution from the contact resistance.
The increasing $T(E)$ at higher binding energies is due to the
opening of new conduction channels. The asymmmetry in the transmission T(E) is 
a function of the NT-pad coupling (C-Au distance). A longer NT-Au distance 
increases the T(E) above $E_F$, while it decreases it below $E_F$, and vise 
versa. Since the NT-pad geometry is kept fixed in all computations, this 
behavior does not influence the effects induced by NT bending.\\

According to our calculation the contact resistance at $E_F$ is only about 5
k$\Omega$, much smaller than the $\approx$ 1 M$\Omega$ resistance typically
observed in experiments on single-wall NTs \cite{tans,martel}. The dependence
of $T(E)$ and contact resistance at $E_F$ on the Au-NT distance is shown in
the upper-left of Figure 2. We see that $T(E_F)$ remains nearly constant
between 1-2 {\AA}, then decreases exponentially.  For distances appropriate
for van der Waals bonding ($\geq 3$ {\AA}) the contact resistance is already
in the M$\Omega$ range. The above findings suggest that the high NT contact
resistance observed experimentally may, in addition to experimental factors
such as the presence of asperities at the metal-NT interface, be due to the
topology of the contact. In most experiments, the NT is laid on top of the
metal pad. The NT is at nearly the van der Waals distance away from the metal
surface, and given that transport in the NT involves high $k$-states which
decay rapidly perpendicular to the tube axis, the coupling between NT and
metal is expected to be weak \cite{note}. Direct chemical bonding between
metal and the NT, or interaction of the metal with the NT cap states
\cite{kim} should lead to stronger coupling.  In this respect, it has been
found \cite{bachtold} that high energy electron irradiation of the contacts
leads to a drastic reduction of the resistance.  Since the irradiation is
capable of breaking NT C-C bonds it may be that the resulting dangling bonds
lead to a stronger metal-NT coupling.

\begin{figure}[tbh]
\centerline{\includegraphics[width=9cm]{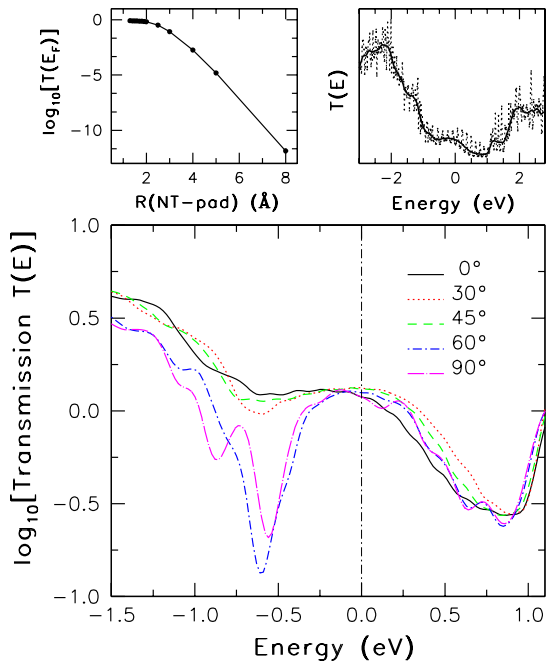}}
\caption{Transmission function $T(E)$ for the different bent nanotubes.
The upper-right inset gives the raw data obtained from the computation for
the straight NT along with the smoothed curve. The upper-left shows the variation 
of the transmission function $T(E_F)$ with the nanotube-pad distance (R(NT-pad)). 
In this last figure, one gold-NT distance is fixed at 1.0 {\AA}
while the other is varied.}
\end{figure}

The strongest modification of $T(E)$ as a result of bending is observed at
around $E$=-0.6~eV where a transmission dip appears. This dip is strongest in
the $60^{\circ}$ bent NT. Furthermore, its transmission function at higher
binding energies (BE) is lower than those of the 0$^{\circ}$-45$^{\circ}$
bent NTs, indicating that the transmission of higher conduction channels is
also decreased. The nature of the dip at about -0.6~eV can be understood by
examining the local density-of-states (LDOS) of bent tubes shown in Figure 3
\cite{rochefort}.  A change (increase) in the LDOS is seen in the same energy
region (0.5-0.8~eV below $E_F$) as the transmission dip. This change is
essentially localized in the vicinity of the deformed region.  The new states
result from the mixing of $\sigma$ and $\pi$ levels, have a more localized
character than pure $\pi$ states leading to a reduction of $T(E)$.  As Figure
3 shows, the change in transmission with bending angle is not gradual; the
transmission of the 30$^{\circ}$ and 45$^{\circ}$ models is only slightly
different from that of the straight tube.  Apparently, large changes in DOS
and $T(E)$ require the formation of a kink in the NT structure, as is the
case in the 60$^{\circ}$ and 90$^{\circ}$ bent NTs.

\begin{figure}[tbh]
\centerline{\includegraphics[width=8cm]{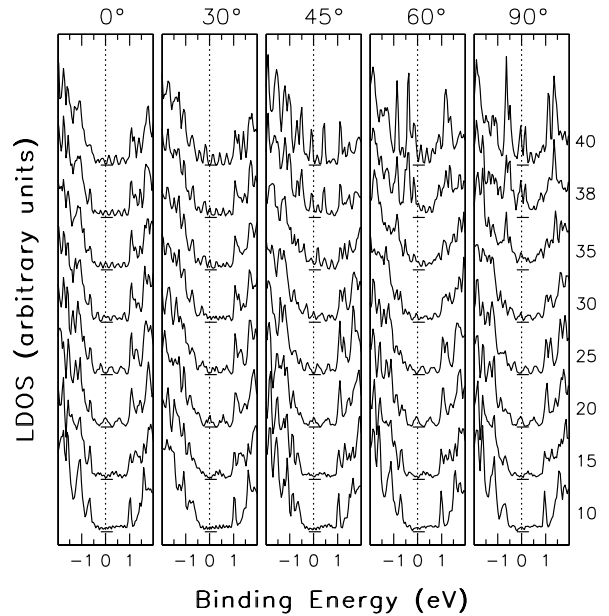}}
\caption{Variation of the LDOS near the Fermi level for several bent 
$(6,6)$ armchair nanotubes. (resolution = 0.05 eV). Indices give the 
relative position of the section in the nanotube structure (1: boundary, 
40: middle of the nanotube)}
\end{figure}

Once the transmission function is computed, the determination of the
differential conductance and resistance is straigtforward.  Figure 4 shows
the results for two extreme cases of equilibration of the Fermi levels. The
first is when $\eta=0$ (Figure 4a), and the symmetric case $\eta=0.5$ (Figure
4b). When $\eta$=0.0, the Fermi level of the NT follows exactly the applied
voltage on one gold pad and the conductance spectrum is directly proportional
to $T(E)$. As expected, there is no large difference between the 0$^{\circ}$,
30$^{\circ}$ and 45$^{\circ}$ models, while the 60 and 90$^{\circ}$ models
show the dip structure at around 0.6~V. The non-linear resistance (NLR)
spectra show clearly a sharp increase by almost an order of magnitude at
0.6~V. These features are also observed when $\eta$=0.5, where the Fermi
level of the NT is floating at half the voltage applied between the two gold
pads. The dip at around 1.2~V in the conductance spectra is now broader,
and the NLR of the 60$^{\circ}$ bent tube increases by about a factor of 4
from the computed resistance of the straight tube. These results suggest that
there exists a critical bending angle (between 45$^{\circ}$ and 60$^{\circ}$)
above which the conduction in armchair carbon nanotubes is drastically
altered \cite{comment}.

\begin{figure}[tbh]
\vskip-0.5cm
\centerline{\includegraphics[width=9cm]{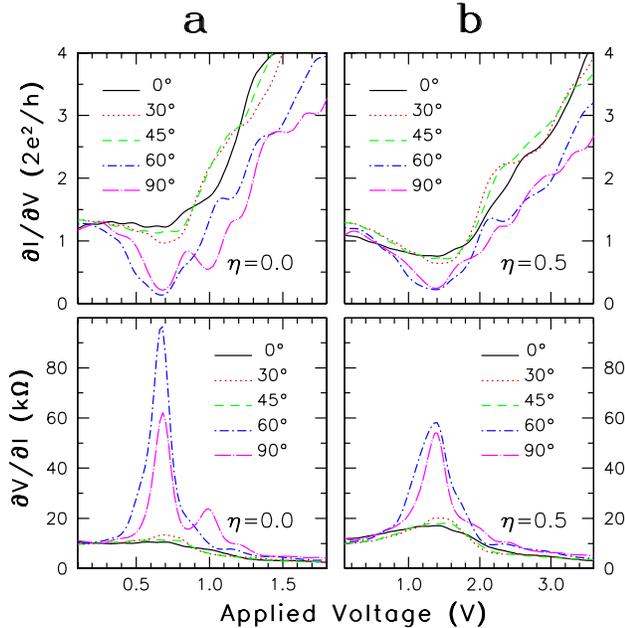}}
\caption{Differential conductance (top) and resistance (bottom) 
of bent tubes for two extreme cases where (a) $\eta$=0.0 and 
(b) $\eta$=0.5}
\end{figure}

In conclusion, we have calculated the effects of structural distortions of
armchair carbon nanotubes on their electrical transport properties. We found
that bending of the nanotubes decreases their transmission function and leads
to an increased electrical resistance. The effect is particularly strong at
bending angles higher than 45$^{\circ}$ degrees when the strain is strong
enough to lead to kinks in the nanotube structure. The electronic structure
calculations show that the reduction in $T(E)$ is correlated with the
presence at the same energy of localized states with significant
$\sigma$-$\pi$ hybridization due to the increased curvature produced by
bending. Resistance peaks near $E_F$ are the likely cause for the experimentally
observed low temperature localization in carbon NTs bent over metal
electrodes \cite{bezryadin}. Our calculations of the resistance (including
the contact resistance) of a perfect NT give a value close to $h/2.4e^2$
instead of $h/4e^2$.  This increase in resistance is solely due to the finite
transmission of the contacts. The much larger contact resistances observed in
many experiments on SWNTs are likely due to the weaker coupling of the NT to
the metal when the NT is simply placed on top of the metal electrodes. We
predict that NTs end-bonded to metal pads will have contact resistances of
only a few k$\Omega$.  Such low contact resistances will greatly improve the
performance of NT-based devices and unmask the Q1D transport properties of
NTs.


\begin{references}
\bibitem{comment} The assumption that the potential drops linearly across the
NT is particularly drastic for highly bent models. In reality, in the highly
deformed region, the voltage would drop in the vicinity of defects.
This voltage drop near defects would decrease the transmission
probability of the propagating electrons, and consequently emphasize the
magnitude of the changes we have computed with our first assumption.\
\end{references}
\end{document}